\documentclass[twocolumn,showpacs,superscriptaddress,preprintnumbers,amsmath,amssymb,prc]{revtex4-2}

\usepackage{graphicx}
\usepackage{dcolumn}
\usepackage{bm}
\usepackage{float}
\usepackage{color}
\usepackage{CJK}
\usepackage{ulem}
\usepackage{array} 
\usepackage{makecell}

\usepackage[colorlinks,linkcolor=blue,urlcolor=blue,citecolor=blue]{hyperref}

\usepackage{tikz,xcolor,hyperref}
\definecolor{lime}{HTML}{A6CE39}
\DeclareRobustCommand{\orcidicon}{
\begin{tikzpicture}
\draw[lime, fill=lime] (0,0) 
circle [radius=0.16] 
node[white] {{\fontfamily{qag}\selectfont \tiny ID}};
\draw[white, fill=white] (-0.0625,0.095) 
circle [radius=0.007];
\end{tikzpicture}
\hspace{-2mm}
}
\foreach \x in {A, ..., Z}{%
\expandafter\xdef\csname orcid\x\endcsname{\noexpand\href{https://orcid.org/\csname orcidauthor\x\endcsname}{\noexpand\orcidicon}}
}

\begin{document}
\begin{CJK*}{UTF8}{gbsn}

\title{Impact of fragment formation on shear viscosity in the nuclear liquid-gas phase transition region}

\author{X. G. Deng (邓先概)\orcidA{}}
\affiliation{Key Laboratory of Nuclear Physics and Ion-beam Application (MOE), Institute of Modern Physics, Fudan University, Shanghai 200433, China}
\affiliation{Shanghai Institute of Applied Physics, Chinese Academy of Sciences, Shanghai 201800, China}
\affiliation{National Superconducting Cyclotron Laboratory, Michigan State University, East
Lansing, MI 48824-1321, USA}

\author{P. Danielewicz\orcidB{}\footnote{Corresponding author: daniel@frib.msu.edu}}
\affiliation{National Superconducting Cyclotron Laboratory, Michigan State University, East
Lansing, MI 48824-1321, USA}
\affiliation{Department of Physics and Astronomy, Michigan State University, East
Lansing, MI 48824-1321, USA}

\author{Y. G. Ma (马余刚)\orcidC{}\footnote{Corresponding author: mayugang@fudan.edu.cn}}
\affiliation{Key Laboratory of Nuclear Physics and Ion-beam Application (MOE), Institute of Modern Physics, Fudan University, Shanghai 200433, China}
\affiliation{Shanghai Research Center for Theoretical Nuclear Physics， NSFC and Fudan University, Shanghai 200438, China}

\author{H. Lin (林豪)}
\affiliation{National Superconducting Cyclotron Laboratory, Michigan State University, East
Lansing, MI 48824-1321, USA}
\affiliation{Department of Physics and Astronomy, Michigan State University, East
Lansing, MI 48824-1321, USA}

\author{Y. X. Zhang (张英逊)\orcidE{}}
\affiliation{China Institute of Atomic Energy, Beijing 102413, China}

\date{\today}

\begin{abstract}

Within the improved quantum molecular dynamic (ImQMD) model we follow the evolution of nuclear matter for planar Couette flow in a periodic box.  We focus on the region of liquid-gas phase transition and extract the shear viscosity coefficient from the local stress tensor, directly following viscosity definition.  By switching on and off the mean field and thus inducing the phase transition, we are able to observe the impact of clumping in the phase-transition region onto the viscosity.

\end{abstract}


\maketitle

\section{Introduction}
\label{introduction}
Shear viscosity is not only one of the crucial bulk properties of the liquids and gases in our surroundings, but also it is important for the study of strongly interacting matter, such as quark gluon plasma (QGP) in early Universe and matter inside neutron stars \cite{Heinz,Teaney,Csernai,Song,Song_CPL,Shen,nstar1,nstar2}. The strongly interacting matter can also undergo  the liquid-gas (LG) phase transitions in lower energy  heavy-ion collisions
~\cite{JP95,Sau76,Dan79,YGM98,YGM05,XFL17,CPL1,GB18,SO21,BB19,JBN,WangR1,WangR2}.  Compilation of data for different substances~\cite{Csernai} demonstrates a drop of the shear viscosity in the region of a phase transition, when the entropy density is employed as a universal reference.  One reason for the drop could be long-range correlations developing in the transition region.  When scaled with entropy density, the reduced shear viscosity of strongly interacting matter appears particularly low in the dense stage of ultra-relativistic collisions~\cite{Csernai,Song,dan85,JLN11,nov14,Shen}, which is often tied to the proximity of the system to the QGP phase transition.

In the hadronic regime, the shear viscosity of nuclear matter around normal density was recently inferred from $\gamma$ decay of the isovector giant dipole resonance populated in a fusion-evaporation reaction~\cite{DB17,Guo}.  In intermediate energy heavy-ion collisions, shear viscosity for nuclear matter was assessed~\cite{barker19} by interpreting stopping data in terms of the Boltzmann-Uhlenbeck-Uehling (BUU) equation~\cite{PD84,LS03}, rooted in the Landau theory, with adjustments in the BUU pertaining to supranormal densities and moderate temperatures.  Viscosity in the LG region was addressed with a combination of Maxwell construction and relaxation time approximation in~\cite{XJ13}.  Quantum Molecular Dynamics (QMD) models have also been used~\cite{DQF14,CLZ13,LiuHL} to provide background circumstances of intermediate-energy collisions, in correlating entropy density and shear viscosity there, with the viscosity estimated either in terms of the mean free path or in kinetic theory~\cite{PD84}.   

In the previous investigations, the peculiar impact of the LG phase-transition, of growing long-range correlations, on the shear viscosity, was never explored.  In the BUU approach, uniform density is imposed and momentum is transported by  moving of nucleons independently~\cite{PD84,LS03,LiBA,LiSX}.  In the Maxwell construction, the shear viscosity coefficient in the phase coexistence region is a linear combination of the coefficients for the two uniform phases~\cite{XJ13}, higher at a given density and temperature, than without the transition.  In this work, we set out to the determine the shear viscosity in the LG phase transition region, while accounting for correlations that underlie the fragment formation in low-density matter.  We employ the so-called an Improved QMD (ImQMD) model~\cite{ZYX06,WN16}, which provides a natural framework for the study of correlations.  We enclose a neutron-proton symmetric system in a periodic box, establish a planar Couette flow inside, and determine the shear viscosity coefficient from the momentum flux across the box, following the definition of the coefficient. This approach is then similar to that in experiments and it works without equilibrium assumptions. Nonetheless, QMD-type models struggle, more than the  BUU-type, in enforcing Pauli principle for nucleon-nucleon ($N$-$N$) collisions~\cite{ZYX18}.  This can lead to undesirable effects in long-term stationary calculations of transport coefficients, as the latter are quite sensitive to Pauli blocking.  Benefiting from the proximity of the system to equilibrium, we know though the blocking factors better than they can be estimated in a QMD model and we exploit that knowledge to circumvent the difficulty in the ImQMD model.

The present article is organized as follows: In Sec.\ \ref{ThreeMethods}, we introduce our simulation model and the analysis framework for extracting the viscosity. In Sec.\ \ref{resultsAA}, we present results for the shear viscosity coefficient of nuclear matter, obtain with and without the mean field interactions turned on, and extract using the SLLOD algorithm which will be explained in next section.  Without the mean field interactions, the LG phase transition disappears.  Sec.\ \ref{resultsAB} supplements the results of \ref{resultsAA} with more discussion of the LG phase-transition and with analysis of the correlations giving rise to the fragments.  Sec.\ \ref{summary} is dedicated to conclusions.

\section{Nuclear system and analysis method}
\label{ThreeMethods}

\subsection{ImQMD model}
\label{ImQMDModel}
Two types of models are employed in practice to simulate heavy-ion collisions and to extrapolate conclusions to the limit of infinite nuclear matter.  One, BUU type~\cite{GFB88}, solves directly the BUU equation.  Another, QMD type~\cite{JA91,XJ16,WN16}, follows molecular dynamics that incorporates elements of the BUU equation.  The BUU models follow nucleon single-particle distributions only, while the QMD models include some multi-nucleon correlations and are thus better suited for the goals of the present work. From the latter type, we choose  the improved quantum molecular dynamic (ImQMD) model~\cite{ZYX06,WN16} for our calculations.

In the ImQMD, we employ a potential energy density~\cite{ZYX06,WN16}, with the spin-orbit term dropped, specifically,
\begin{equation}
\begin{split}
V_{loc}&= \frac{\alpha}{2}\frac{\rho^{2}}{\rho_{0}} + \frac{\beta}{\gamma+1}\frac{\rho^{\gamma+1}}{\rho_{0}^{\gamma}} +\frac{g_{sur}}{2\rho_{0}} \, (\bigtriangledown\rho)^{2}     \\
           &+g_{\tau} \frac{\rho^{\eta +1}}{\rho_{0}^{\eta}} +\frac{g_{sur,iso}}{\rho_{0}}\,[\nabla(\rho_{n}-\rho_{p})]^2+\frac{C_s}{2\rho_{0}}\rho^{2} \, \delta^{2} \,,
\end{split}                              
\label{QMDpotential}
\end{equation}
where $\rho = \rho_\text{n} + \rho_\text{p}$ is the net nucleon density, $\rho_\text{n}$ and $\rho_\text{p}$ are the proton and neutron densities, respectively, and $\delta = (\rho_{n}-\rho_{p})/\rho$ is the normalized asymmetry of neutron and proton.  The normal density is $\rho_{0} = 0.16~  \text{fm}^{-3}$.

The Skyrme density functional with interactions such as Eq.~\eqref{QMDpotential}, as well as other approaches to nuclear systems that aim at a level of realism, yield the aforementioned liquid-gas phase transition at the subnormal densities for the near-symmetric nuclear matter.  In the model of nuclear-matter system, the Coulomb interactions are switched off and the near symmetric matter has comparable numbers of neutrons and protons. But in this work, we also check the effect of Coulomb interactions. When uniform fermionic liquid and gas phases are assumed, they may coexist in equilibrium with each other.  At the ground state ($T$ = 0), the two phases are nuclear matter at normal density and vacuum.  Difference in the densities of the model phases in equilibrium shrinks as the temperature increases and the difference disappears as both phases reach critical density and temperature, 
$\rho_c \sim 0.4 \, \rho_0$ and $T_c  \sim 18 \, \text{MeV}$  \cite{Sau76,Dan79}, respectively. Since the fluctuations exist in the non-equilibrium nuclear system, the phase transition picture becomes more complex, with fragments forming at lower temperatures in the phase-transition region and the fragment formation probability representing a $U$-shape as a function of fragment mass, e.g., Ref.~\cite{Fu06}.  The light-mass arm of the $U$-shape represents the gas phase and the heavy-mass arm -- the liquid phase.

With our focus on the shear viscosity of infinite symmetric nuclear matter, we set up a periodic cubic box as in Ref.~\cite{ZYX18}.  In our simulations, we use a fixed number of nucleons $A = 600$.  The box size is adjusted to yield a desired average density. The width of the Gaussian wave-packets that represent the nucleons, which has some impacts on the fragments that are formed in dilute matter, is set at $\sigma = 2\,\text{fm}$. For simplicity, the $N$-$N$ cross section is fixed at $40 \,\text{mb}$ in this paper. Even though the value of shear viscosity  increases with the decreasing of  $N$-$N$ cross section as shown in our previous work \cite{CLZ13} as well as our check in this work, the choice of  $40 \,\text{mb}$ for the $N$-$N$ cross section does not change our conclusion of the present work . In QMD models, the Pauli blocking for $N$-$N$ collisions is normally calculated according to the wave-packets, with the blocking fluctuating at low temperatures and the momentum distributions gradually evolving towards Boltzmann, rather than Fermi-Dirac  form~\cite{ZYX18}.  Benefiting from the proximity of our systems to a local equilibrium, we circumvent the problem using the local equilibrium Fermi-Dirac distributions, corresponding to the local nucleon and kinetic energy density for which  collective motion is deducted, in the blocking for the collisions.  A similar procedure was employed in the transport-code comparison~\cite{ZYX18}.  And in the following, the temperature is an input parameter of the ImQMD model, and initial momentum distributions of the nucleon is  determined by the Fermi-Dirac distribution at finite temperature as performed in our earlier Thermal IQMD (ThIQMD) model \cite{DQF14,WangTT}.

In this work, we consider both of cases without and with mean field. In the case without mean field and in the cascade mode, the temperature of the system would not change during the process of simulation. However, when one turns on the mean field, the nucleon potential and kinetic energy will change due to the fragment formation. Then the system would heat up, and we need to adjust this transient state temperature to the initial temperature by introducing the factor `$h$' which can be nominally positive and less than `1'. Here the transient state temperature is determined  by fitting the momentum distribution with Fermi-Dirac function at certain times. If one can find that the transient state temperature increases as time, we need  a friction factor `$h$' to cool the system. Also the `$h$' is only implant in a period of time (the length of time depends on the initial density and temperature). One needs to adjust it to be an appropriate value, eg. `$h$' = 0.999 for density of 0.1$\rho_{0}$ and $T$ = 6 MeV. In practice, the `$h$' is implemented when the box system is close to the equilibrium, and then the box system cools down to initial temperature and reaches equilibrium again after some time evolution. One problem for our calculations is that the term can lead to phase space occupations inconsistent with the Pauli principle for fermions. We resolve that by employing the term only at high momenta and/or low spatial densities, under the conditions of low phase-space density.

\subsection{Gaussian thermostated SLLOD algorithm}
\label{GTSLLOD}

The SLLOD algorithm belongs to common methodologies of the Non-Equilibrium Molecular Dynamics (NEMD) calculations. It  was named by Evans and  Morriss \cite{GP06} and related to the dynamics with (artificial) strain rate  algorithm  for non-equilibrium molecular dynamics  calculation and has been  extensively applied to predict the rheological properties of real fluids \cite{GP06,DJE08}.  

In the SLLOD algorithm, a planar Couette flow field is applied to the system, with a shear rate $\dot{\gamma}$ = $\partial v_{x}/\partial y$.  When the shear is presented, the periodic boundary conditions are modified to those of Lees and Edwards~\cite{LE72}, consistent with constant shear throughout space. The shear viscosity coefficient is computed from \cite{DJE08}
\begin{equation}
\begin{split}
\eta = -\frac{\langle P_{xy} \rangle }{\dot{\gamma}} \, ,
\end{split}                              
\label{GKubo1}
\end{equation}
where $\langle\cdots\rangle$ denotes ensemble average and the stress tensor is given by \cite{RG09}
\begin{equation}
P_{\alpha \beta} = P_{\alpha \beta}^{\rm cont}+ P_{\alpha \beta}^{\rm coll}\,.
\end{equation}
There are two contributions to the stress tensor, the continuous one and the collision impulse one. The continuous contribution to the stress tensor is given by
\begin{equation}
\begin{split}
P_{\alpha\beta}^{\rm cont}(t) =\frac{1}{V}\int {\rm d}^{3}r \,  P_{\alpha\beta}(\vec{r}, t) \; \; \; (\alpha, \beta=x,y,z) \, .
\end{split}                              
\label{GKubo3}
\end{equation}
Here, $V$ is the volume of the simulation box and  $P_{\alpha\beta}(\vec{r}, t)$ originates from summing over the wave packets and is formally defined as \cite{Deng2021}
\begin{equation}
\begin{split}
&P_{\alpha\beta}(\vec{r},t)= \sum_{i}^{A} \frac{p_{{i}\alpha} \, p_{{i}\beta}}{m_{i}} \, \rho_{i}(\vec{r},t)   \\
&+\frac{1}{2} \sum_{i}^{A} \sum_{i\neq j}^{A} F_{ij\alpha} \, R_{ij\beta} \, \rho_{j}(\vec{r},t)             \\   
 &+\frac{1}{6} \sum_{i}^{A} \sum_{i\neq j}^{A}\sum_{i\neq j \neq k}^{A} (F_{ijk\alpha} \, R_{ik\beta}+ F_{jik\alpha} \, R_{jk\beta} )\rho_{k}(\vec{r},t) \\
&+\cdots  \, .   
\end{split}                              
\label{GKubo4}
\end{equation}
Here, $\vec{F}$ are two- and three-body forces that are continuous as a function of time.  The Latin indices pertain to the particles and the first index in the force represents the particle on which the force acts and the remaining particle indices indicate the particles that give rise to that force.  The relative positions of two particles are 
$\vec{R}_{ij} = \vec{r}_{i}-\vec{r}_{j}$. In QMD-type models, contribution to the density associated with particle $i$ can be written in terms of a wave packet as
\begin{equation}
\begin{split}
\rho_{i}(\vec{r}) = \frac{1}{(2\pi \sigma^{2})^{3/2}} \exp \Big[-\frac{(\vec{r}-\vec{r}_{i})^{2}}{2\sigma^{2}} \Big]
\end{split}                              
\label{GKubo5}
\end{equation}
During collisions between nucleons, momentum is transported through action of impulse forces.  All collisions within time interval $\Delta t$ yield a time-averaged contribution to the stress tensor equal to \cite{RG09}
\begin{equation}
\begin{split}
P_{\alpha\beta}^{coll}(t) = \frac{1}{V}\sum_{i}^{Ncoll} \frac{\Delta p_{i 1\alpha }}{\Delta t}\cdot(r_{i1\beta}-r_{i2\beta}) \,.
\end{split}                              
\label{GKubo6}
\end{equation}
Here, the summation is over nucleon-nucleon collisions, with the two nucleons in the collision labelled as '1' and '2'.  The factor ${\Delta p_{i1\alpha}}$ is the $\alpha$ component of the impulse on nucleon 1 in the collision.  The factor $(r_{i1\beta}-r_{i2\beta}) $ is the $\beta$ component of the displacement vector over which the momentum is moved in the collision. 

\begin{figure}[htb]
\setlength{\abovecaptionskip}{0pt}
\setlength{\belowcaptionskip}{8pt}
\includegraphics[scale=1.10]{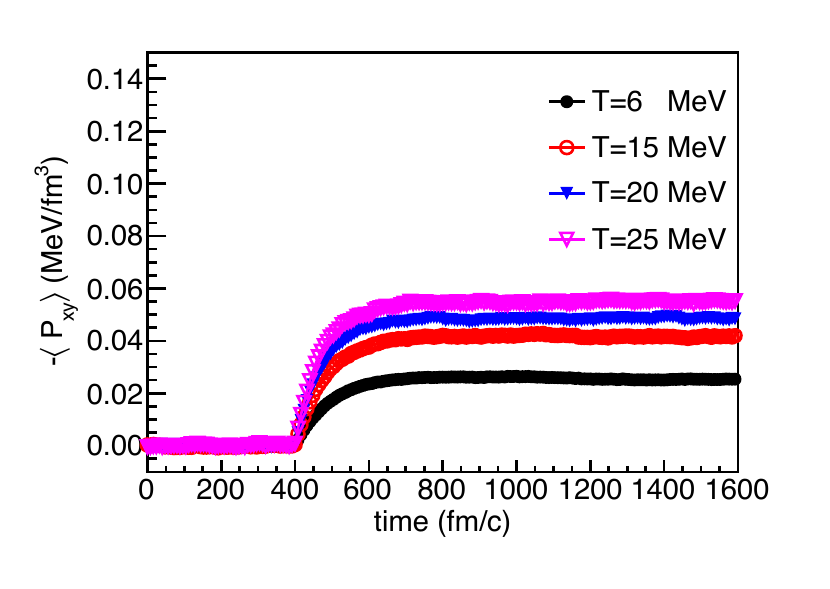}
\caption{The $xy$ element of the stress tensor, as function of time at the density of $0.1 \, \rho_{0}$, for different temperatures in the simulations.}
\label{fig:fig1}
\end{figure}

\begin{figure}[htb]
\setlength{\abovecaptionskip}{0pt}
\setlength{\belowcaptionskip}{8pt}
\includegraphics[scale=1.08]{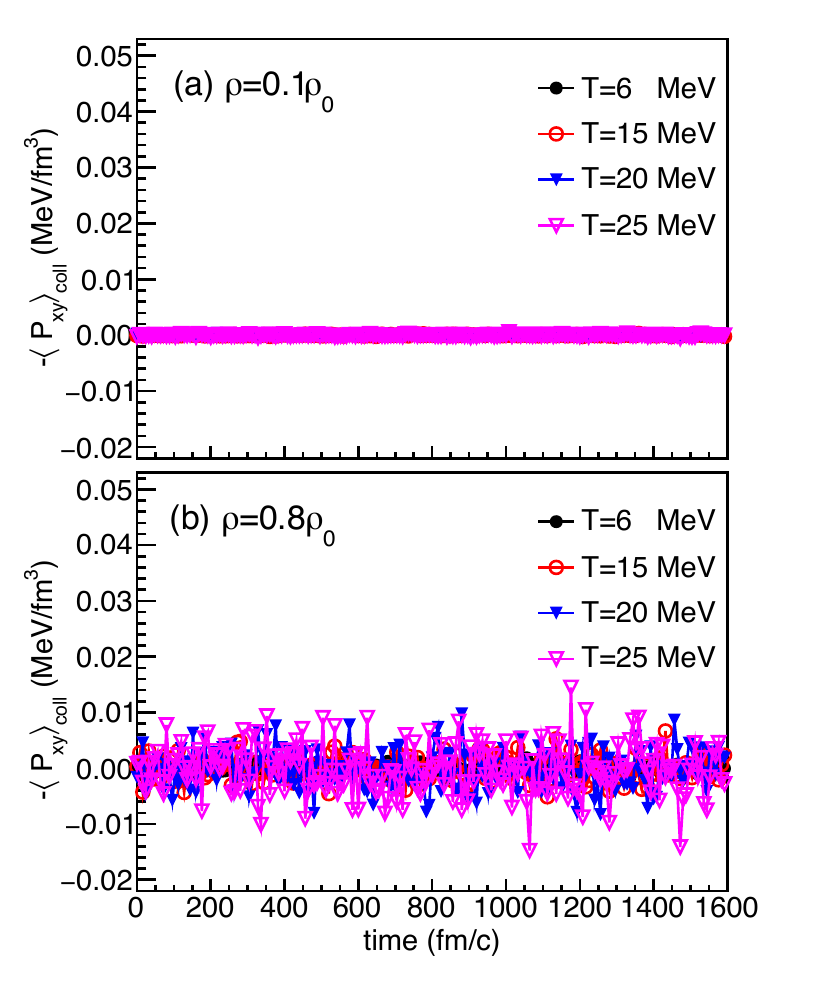}
\caption{Collision contribution to the off-diagonal stress tensor element, as function of time at the densities of $0.1 \, \rho_{0}$ (a)  and $0.8 \, \rho_{0}$ (b), for different temperatures in the simulations.}
\label{fig:fig2}
\end{figure}

Fig.~\ref{fig:fig1} shows the evolution of the off-diagonal element of the stress tensor for the system, first allowing the system to equilibrate during the first $400 \, \text{fm}/c$ and then applying the shear of the Couette flow within the box, still to be discussed.  Without shear, the element fluctuates around zero.  After the shear is switched on, the element quickly rises and stabilizes after about $100 \, \text{fm}/c$.  The collision contribution to the element is very small, as illustrated for different densities in Fig.~\ref{fig:fig2}.  The fluctuations are stronger at the higher density, but still very much near zero.

\begin{figure}[htb]
\setlength{\abovecaptionskip}{0pt}
\setlength{\belowcaptionskip}{8pt}
\includegraphics[scale=1.1]{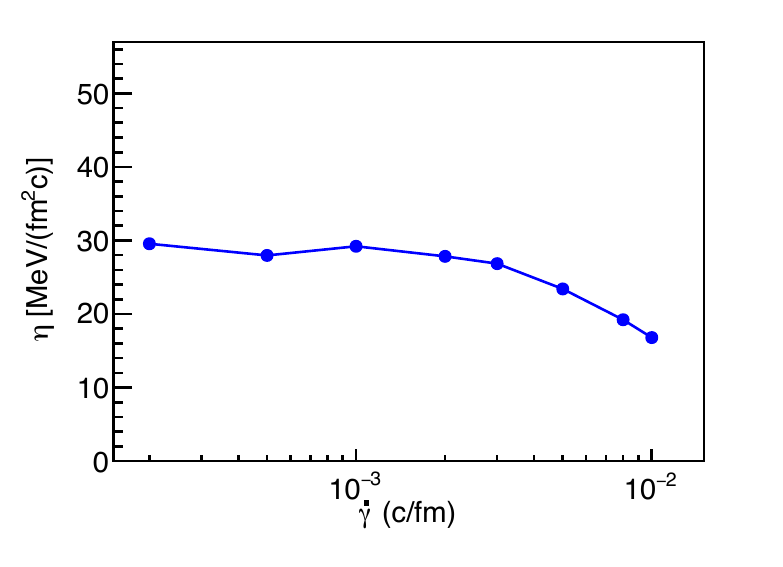}
\caption{ Deduced shear viscosity as function of the shear rate $\dot{\gamma}$ applied within the SLLOD algorithm, in the system at the density of $0.2 \, \rho_{0}$ and temperature $T = 10 \, \text{MeV}$, when no mean field is employed.}
\label{fig:fig3}
\end{figure}

With the applied planar Couette flow field, the centroids of nucleonic wave-packets satisfy the SLLOD equations \cite{DJE08}:
\begin{align}
&\frac{d\vec{r}_{i}}{dt}=\frac{\vec{p}_{i}}{m_{i}} + \dot{\gamma}y_{i} \hat{x}       \, ,        
\label{GKubo77}  \\
&\frac{d\vec{p}_{i}}{dt}=\vec{F}_{i}-\dot{\gamma}p_{yi} \hat{x}\, .
\label{GKubo88}
\end{align}
The construction is that of introducing a local velocity with the component in the $x$ direction changing linearly in the $y$ direction.  The $\dot{\gamma}$ factor regulates the pace of the velocity change in the $y$ direction. This, together with the Lees-Edwards periodic boundary condition, allows to combine a uniform shear with periodicity.

One should notice that the shear rate $\dot{\gamma}$ can neither be too weak, nor too strong~\cite{GP05,CD08}.  If it is too weak, the induced off-diagonal  stress tensor elements will have hard time competing with numerical inaccuracies.  If it is too strong, the response of the system will cease to be linear in the shear.  In either case, the extraction of the shear viscosity is likely to fail.  In Fig.~\ref{fig:fig3} we show  the extracted viscosity as a function of the shear rate in an exemplary system at the density of $0.2 \, \rho_0$ and temperature $T = 10 \, \text{MeV}$, when no mean field is applied.  In most of the calculations, we employ the rate of $\dot{\gamma}$ = 0.002 $c$/fm when the system reaches equilibrium.

\begin{figure}
\setlength{\abovecaptionskip}{0pt}
\setlength{\belowcaptionskip}{8pt}
\includegraphics[scale=1.1]{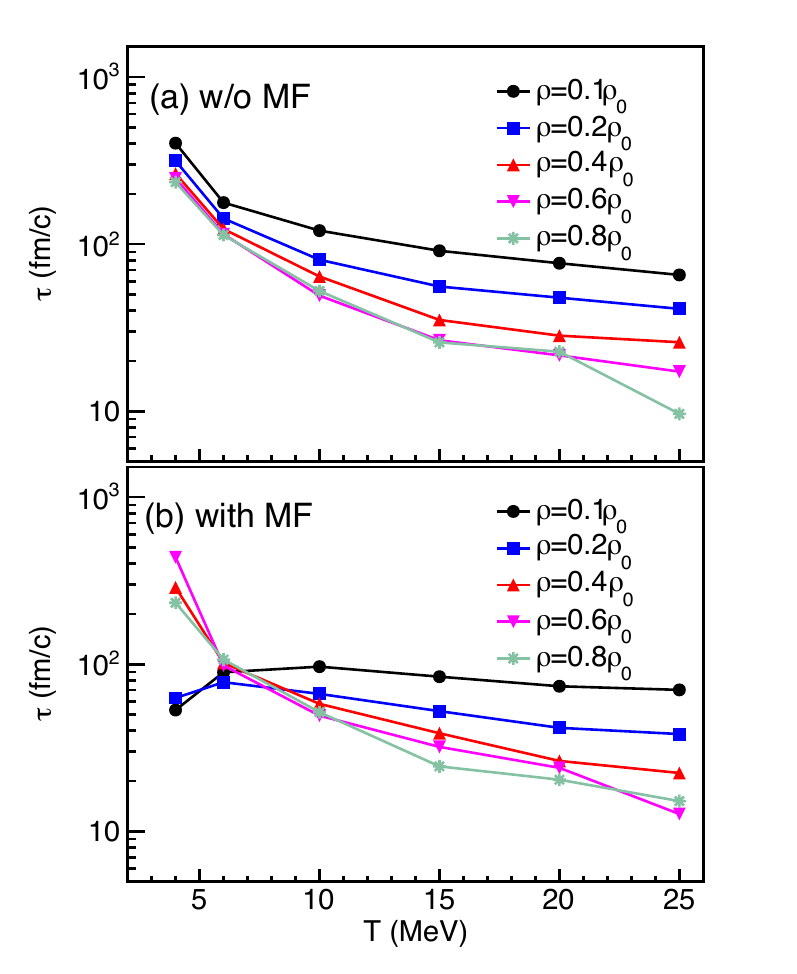}
\caption{Relaxation time as function of temperature at different densities in the calculations without (a) and with (b) mean field.} 
\label{fig:fig4-0}
\end{figure}

\begin{figure}
\setlength{\abovecaptionskip}{0pt}
\setlength{\belowcaptionskip}{8pt}
\includegraphics[scale=1.1]{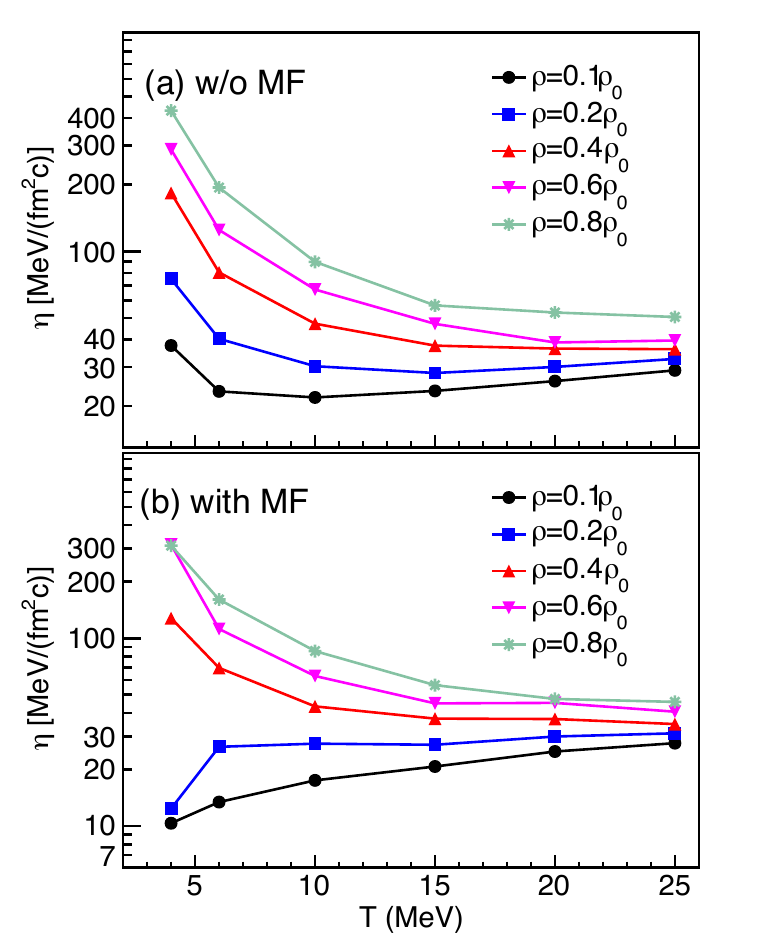}
\caption{Same as Fig.~\ref{fig:fig4-0} but for shear viscosity.}
\label{fig:fig4}
\end{figure}

\begin{figure}
\setlength{\abovecaptionskip}{0pt}
\setlength{\belowcaptionskip}{8pt}
\includegraphics[scale=1.1]{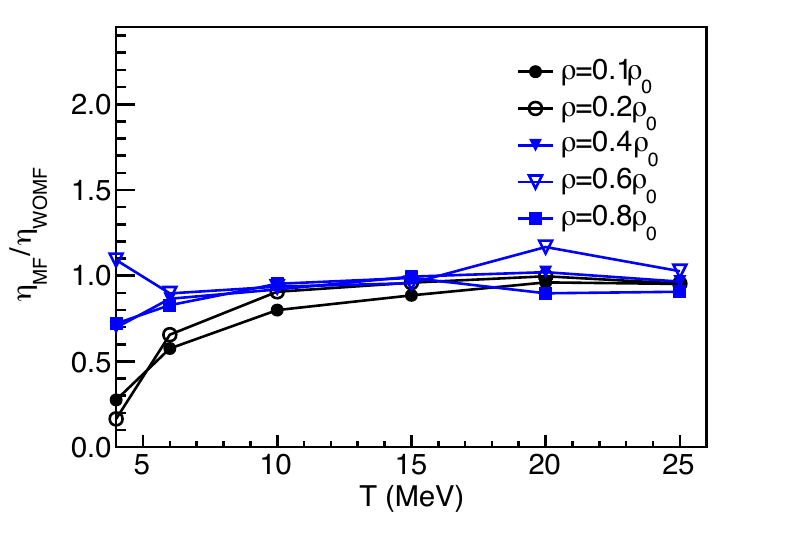}
\caption{The ratio of the shear viscosity with mean field to that without as  function of temperature  at different average density.}
\label{fig:fig5}
\end{figure}

\section{Results and discussion}
\label{resultsSH}

\subsection{Impact of fragment formation on shear viscosity}
\label{resultsAA}

For a viscoelastic system, relaxation time is required for the establishment of the stationary state. For our simulations, shear rate $\dot{\gamma}$ is implemented into the system at 400 fm/c as shown in Fig.~\ref{fig:fig1}. Stress tensor $P_{xy}$ responds to the shear rate and $-\langle P_{xy} \rangle$ increases as time  and reaches constant for some times. For  responding to  shear rate $\dot{\gamma}$ (t) which actually is set to be constant, stress tensor $P_{xy}(t)$ can be expressed as \cite{DJE08}
\begin{equation}
\begin{split}
P_{xy}(t) = - \int_{t_{0}}^{t} \eta_{M} (t-s) \dot{\gamma} (s) ds \, ,
\end{split}                              
\label{MFunction1}
\end{equation}
where $\eta_{M} (t)$ is a memory function and $t_{0}$ is the time for implementation of shear rate. For the Maxwell model of viscoelasticity, the memory function is identified as \cite{DJE08, HM78,RH13} 
\begin{equation}
\begin{split}
\eta_{M} (t) = G\exp (-t/\tau) \,,
\end{split}                              
\label{MFunction2}
\end{equation}
where $G$ is the infinite frequency shear modulus and $\tau$ is relaxation time. With Eq. \ref{MFunction1} and Eq. \ref{MFunction2}, one can get simple expression for the stress tensor:
\begin{equation}
\begin{split}
P_{xy}(t) = A + B \exp (-t/\tau)\,.
\end{split}                              
\label{MFunction3}
\end{equation}
By fitting the curves as in Fig.~\ref{fig:fig1} with Eq. \ref{MFunction3}, one can get the relaxation time which is shown in Fig.~\ref{fig:fig4-0}. Fig.~\ref{fig:fig4-0} (a) and (b) display a decreasing of  relaxation time with the  increasing of temperature. In Fig.~\ref{fig:fig4-0} (b), it is found  that at low $T$ and density of low energy nuclear LG phase transition domain the relaxation time is reduced. For the hadron gas or hadron mixture as shown in Refs. \cite{MP93,AM04,JB18}, the relaxation time has the same order of magnitude but within  different temperature region.

In Fig.~\ref{fig:fig4} we show our major results, i.e., the shear viscosity coefficient calculated for symmetric nuclear matter at different densities, using the SLLOD algorithm, with and without mean field.  Our treatment of the Pauli principle allows us to push the calculations down to relatively low temperatures, but it does not eliminate completely the problems around immediate vicinity of zero temperature and we restrict ourselves to $T \ge 4 \, \text{MeV}$.

Without mean field, the viscosity is that of a gas of nucleons.  Comparing with the results elsewhere, ours include collisional contributions, but these are small in the density region we consider, cf.~Fig.~\ref{fig:fig2}.  Because of this, our results in Fig.~\ref{fig:fig4}(a), obtained without MF, are similar to those elsewhere, such as in Ref.~\cite{PD84}. In low temperature region, the nucleonic gas becomes degenerate and the Pauli blocking begins to play a role for collisions, the shear viscosity begins to increase with the decreasing of temperature in the degenerate gas. The latter is due to the increase in the mean free path because collisions become increasingly blocked as temperature decreases. And in high temperature region, the purely nucleonic shear viscosity coefficient slowly increases with temperature due to the increase in nucleon velocities with temperature.

\begin{figure}
\setlength{\abovecaptionskip}{0pt}
\setlength{\belowcaptionskip}{8pt}
\includegraphics[scale=1.1]{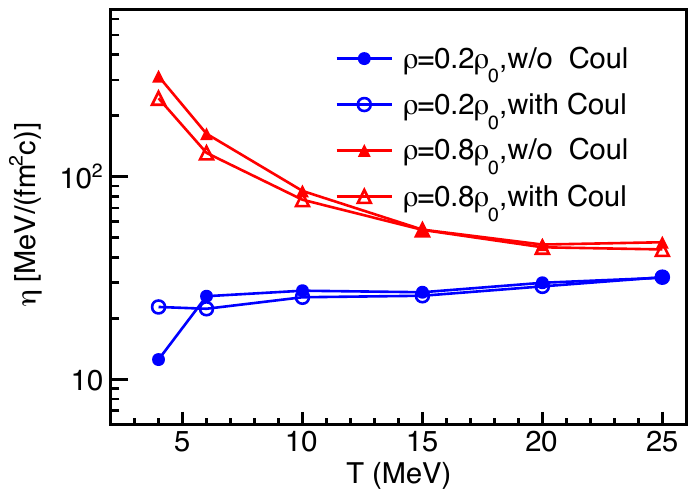}
\caption{Shear viscosity as function of temperature in the calculations without and with the Coulomb interaction.}
\label{fig:fig6}
\end{figure}

\begin{figure}
\setlength{\abovecaptionskip}{0pt}
\setlength{\belowcaptionskip}{8pt}
\includegraphics[scale=1.1]{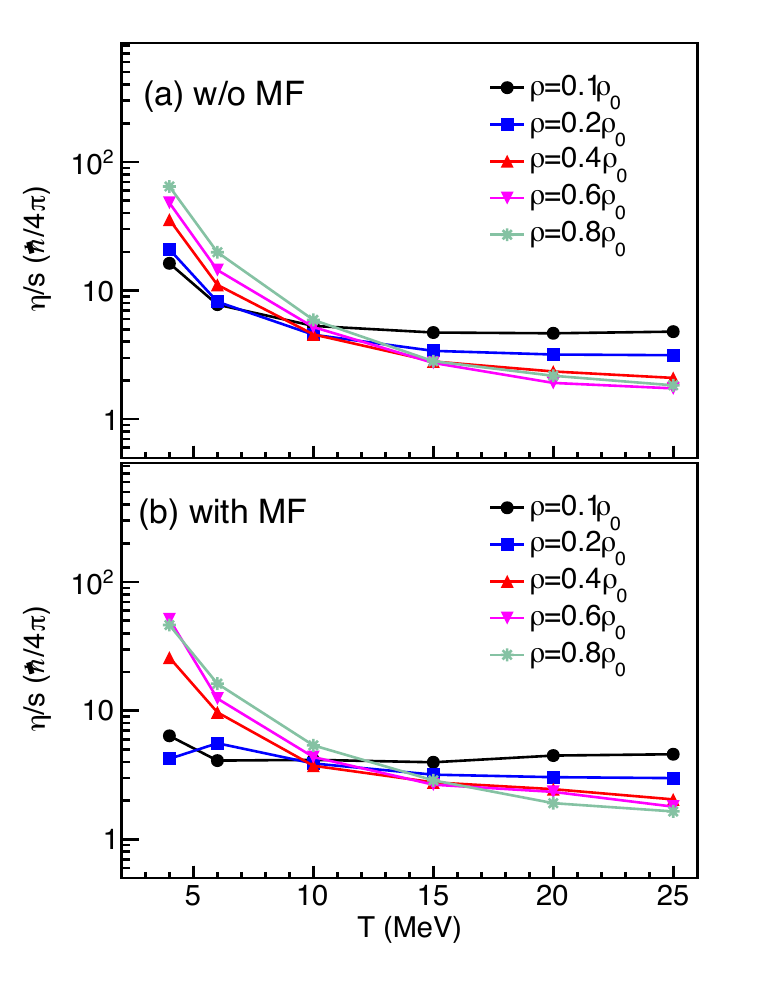}
\caption{Same as Fig.~\ref{fig:fig4-0} but for the ratio of shear viscosity over entropy density ($\eta/s$). }
\label{fig:fig6-0}
\end{figure}
With MF included, our results for shear viscosity in Fig.~\ref{fig:fig4}(b) are similar to those without MF at higher temperatures and densities. 
To facilitate the comparison, we show in Fig.~\ref{fig:fig5} the ratio of the shear viscosity obtained with mean field over that obtained without,  as a function of temperature at given average densities.  The ratio is close to unity at temperatures in excess of $\sim 12 \, \text{MeV}$ and at densities closer to the normal density $\rho_0$.  At low densities, below $0.4 \, \rho_0$ and temperatures less than $10 \, \text{MeV}$, we observe a strong reduction in the shear viscosity when the mean field is included.  
In our system, a strong reduction in the shear viscosity  occurs in the region of the phase transition and can be understood in terms of formation of separated fragments that move slowly, absorb most of the mass in the system, and stall the transport of momentum.  By moving slowly, the fragments play only a passive role in transporting the momentum.  The latter gets primarily transported by the nucleons in the gas phase, but these rather collide with the fragments than with each other, which shortens their mean free path.  In addition, with most mass in the fragments, the role of the Pauli principle in the gas is reduced, additionally reducing the path. At higher densities, the fragments connect and the low density gas phase is contained to voids in the liquid, so momentum transport can progress.  Any role of the Pauli principle is restored. Moreover, while the fragments forming, the long-range Coulomb interaction  may have effect  on shear viscosity. For checking, in Fig.~\ref{fig:fig6}, Coulomb interaction is taken into account. It is found that in the region of low temperature, shear viscosity is reduced by the Coulomb interaction. Imagining two fluid layers passing from one to another, Coulomb interaction of among protons which is repulsive will reduce momentum  transformation or momentum flux between two layers that will decrease shear viscosity. In low temperature region, 
Coulomb interaction could be important in comparison with thermal motion. While with increasing temperature, Coulomb interaction effect becomes smaller since nuclear matter at higher temperature is more uniform and the weight of Coulomb interaction becomes negligible. Based on the above arguments, we can explain the temperature dependent Coulomb correction on shear viscosity.

It is interesting to note that for the quark gluon plasma by taking into account interaction, shear
viscosity faces a strong suppression in low temperature region of quark hadron phase transition domain  going from $T$ = 0.300 GeV to 0.120 GeV in contrast with the noninteracting QGP case \cite{IJMPE}. Our results for the reduction of shear viscosity are quite similar to this QGP  phase transition even though our system is for LG system.

For a simplified picture, by utilizing the relation \cite{KK85}, 
\begin{equation}
\begin{split}
s = \frac{g}{(2\pi\hbar)^{3}V}  \int [f\ln f+(1-f)\ln (1-f)]d^{3}p d^{3}r\, ,
\end{split}                              
\label{MFunction4}
\end{equation}
one can get the entropy density ($s$), where $g = 4$ is spin-isospin degeneracy and  $f$ is the distribution function, which can be given,
\begin{equation}
\begin{split}
f(\epsilon) = \frac{1}{\exp(\frac{\epsilon(\vec{r})-\mu(\vec{r})}{T})+1}
\end{split}                              
\label{FermiDiracDis}
\end{equation}
where $\mu$ is chemical potential,  and the particle energy $\epsilon$ = $p^{2}/(2m)+$$U(\rho(\vec{r}))$ in which $U(\rho(\vec{r}))$ represents single particle potential. Here, by taking into account fragment effects in Eq.~(\ref{FermiDiracDis}), we can obtain local entropy densities with the local chemical potentials. After  integration local entropy densities over the whole space, the total entropy density can be got.

\begin{figure}[htb]
\setlength{\abovecaptionskip}{0pt}
\setlength{\belowcaptionskip}{8pt}
\includegraphics[scale=1.1]{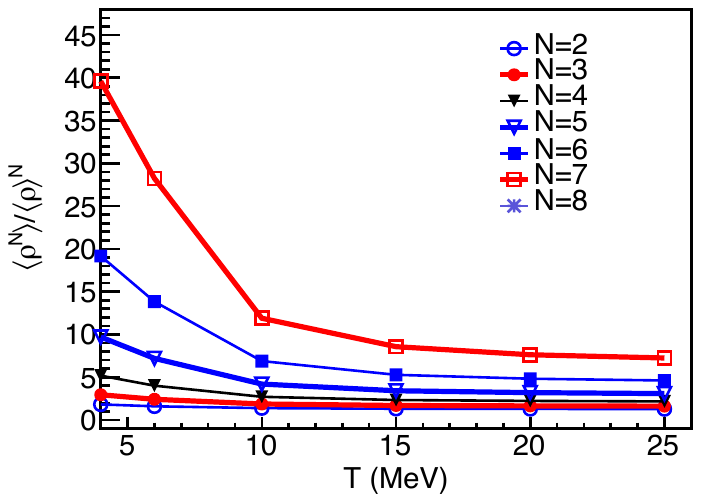}
\caption{Reduced density moments for different powers $N$ vs temperature at the average density of 0.2$\rho_{0}$ from the calculations with mean field.}
\label{fig:fig7}
\end{figure}

With shear viscosity we have as in Fig.~\ref{fig:fig4}, the ratio of shear viscosity over entropy density ($\eta/s$) which is a hot agenda and relates to a bound value of $\hbar/4\pi$, known as the Kovtun-Son-Starinets (KSS) bound \cite{KSS05} can be obtained. The hot quark gluon plasma (QGP) has very low $\eta/s$ which is very close to $\hbar/4\pi$ and behaves as a nearly perfect fluid \cite{TSDT09}. When the QGP cools down, in phase of relativistic hadron gas, $\eta/s$ increases and becomes significantly higher than the KSS bound~\cite{KI08,ND09}. For the finite colder nuclear matter around the saturation density, it is 2.5$-$6.5 times of $\hbar/4\pi$ within the temperature range 0.8$-$2.1 MeV \cite{DB17,Guo}. For the hot finite nuclear matter, $\eta/s$ is around 3.0 - 70 times of $\hbar/4\pi$ for the density of 0.2 - 1.25 $\rho_0$ \cite{LiuHL,DQF14,CLZ13}, in which the minimum $\eta/s$  value corresponds to the liquid-gas phase transition in previous model calculations. And here we get $>$ 2 times of $\hbar/4\pi$ for the infinite nuclear matter which is consistent with the results in Ref.~\cite{   CLZ13,LiuHL,LiSX,XJ13,Deng2016}.  To briefly summarize the values of ratios of shear viscosity over entropy density for different systems,  Table \ref{ETAOS1} and Table \ref{ETAOS2} are made.  From tables, we can see the QGP matter around 170 MeV has the lowest $\eta/s$ close to the KSS bound ($1/4\pi$), however, the nuclear matter around a few to a few tens MeV which is in a liquid gas phase transition range has also relative low $\eta/s$, i.e. about several times of KSS bound. The above low $\eta/s$ may reflect the universal property of strong interaction matter regardless of partonic and nucleonic level. In contrary, for atomic and molecular substances of He, Ni and water, they have relative large $\eta/s$ which are dominated by electromagentic interaction.

\begin{table}[htp]
\setlength{\belowcaptionskip}{0.2cm}
\centering  
\caption{$\eta/s$ of different atomic and molecular systems:  Helium, Nitrogen and H$_{2}$O in certain ranges of  temperature and pressure \cite{Csernai}. } 
\label{ETAOS1}
\begin{tabular}{|p{3cm}|p{1.5cm}<{\centering}|p{1.5cm}<{\centering}|p{1.5cm}<{\centering}|}
\hline
Systems [Ref.]  &  T (K)    & Pressure (MPa) & $\eta/s~(\hbar/4\pi)$  \\ %
\hline
Helium \cite{Csernai}  &  2$-$20 & 0.1$-$1 & 8.8$-$126  \\ %
\hline
Nitrogen \cite{Csernai}  & 50$-$ 600& 0.1$-$10 & 11.9$-$7000  \\ %
\hline
H$_{2}$O \cite{Csernai}  &  300$-$1200 & 10$-$100 & 25.5$-$377  \\ %
\hline
\end{tabular}
\end{table}
\begin{table}[htp]
\setlength{\belowcaptionskip}{0.2cm}
\centering  
\caption{$\eta/s$ of different quark matter (QGP) and finite nuclear matter (FNM) as well as infinite nuclear matter (INM)  in certain ranges of  temperature and density. Here T$_{c}$$\approx$ 170 GeV is transition temperature from the hadronic phase to QGP phase \cite{IJMPE}. } \label{ETAOS2}
\begin{tabular}{|p{3cm}|p{1.5cm}<{\centering}|p{1.5cm}<{\centering}|p{1.5cm}<{\centering}|}
\hline
Systems [Refs.]  &  T (MeV)  &  $\rho$/$\rho_{0}$  &  $\eta/s~(\hbar/4\pi)$  \\ %
\hline
QGP1~\cite{IJMPE,CS10,MR13,PDeb16}  &  $\leqslant$$\rm T_{c}$ & --- & 3.77$-$25.1    \\ %
\hline
QGP2 \cite{IJMPE,CS10,MR13,PDeb16,DDD11} &  $>$$\rm T_{c}$ & --- & 1.00$-$6.91    \\ %
\hline
FNM1 \cite{Guo,DB17}  &  0.8$-$2.1  & $\sim$1.0 &   2.5$-$6.5  \\ 
\hline
FNM2 \cite{LiuHL}  &  3.5$-$16  & 0.9$-$1.25 &   3.0$-$70.0  \\ %
\hline
FNM3 \cite{DQF14}  &  1$-$30  & 0.3$-$1.0 &   7.0$-$60.0  \\ %
\hline
FNM4 \cite{CLZ13}  &  6$-$16  & 0.2$-$0.3 &   9.5$-$20.0  \\
\hline
INM1 \cite{XJ13}  &  8$-$14 &  0.01$-$0.3 & 4$-$30 \\ %
\hline
INM2 $[$this work$]$ &  4$-$25 & 0.1$-$0.8 & 2$-$55   \\ %
\hline
\end{tabular}
\end{table}

\begin{figure}[htb]
\setlength{\abovecaptionskip}{0pt}
\setlength{\belowcaptionskip}{8pt}
\includegraphics[scale=1.1]{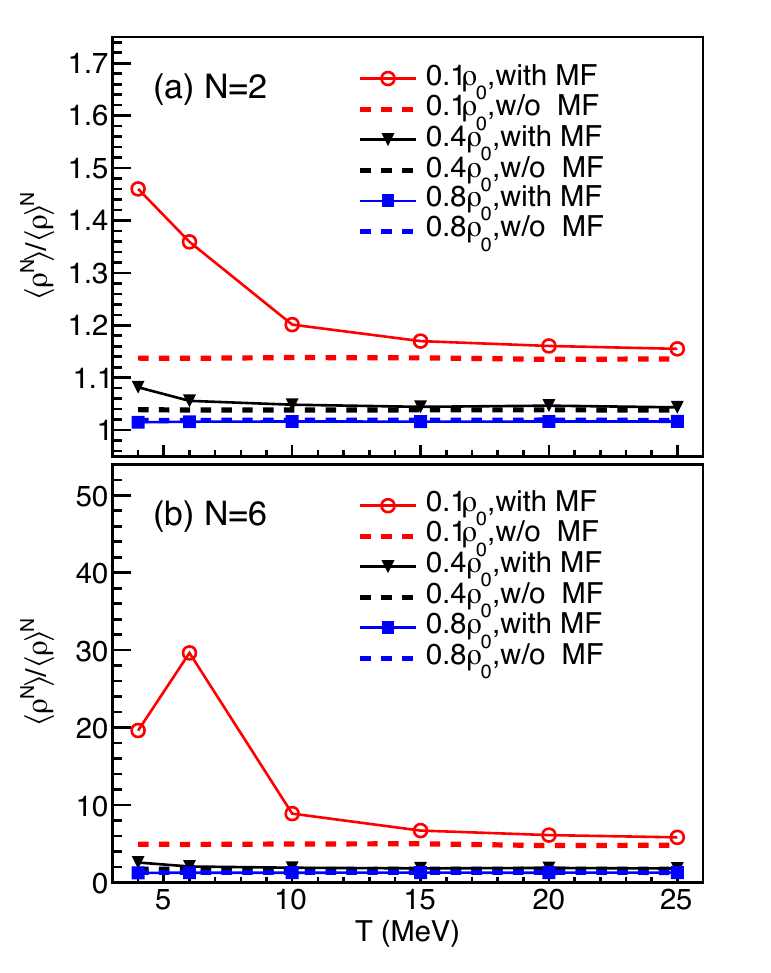}
\caption{Density moment as function of temperature with (the lines with symbols) and without (the dash lines) mean field for different densities with $N$ = 2 (a) and $N$ = 6 (b).}
\label{fig:fig8}
\end{figure}

\subsection{Density moments}
\label{resultsAB}

The LG phase-transition region and fragment formation are characterized by enhanced density fluctuations~\cite{IV08}.  Similar fluctuations would be there for the anticipated QCD phase transition in baryon-rich matter~\cite{CS07}.  The density fluctuation can be quantified in terms of average density moment~\cite{JS12},
\begin{equation}
\begin{split}
\langle \rho^{N} \rangle  = \frac{1}{A} \int \rho(\vec{\bf{r}})^{N} \rho(\vec{\bf{r}}) d^{3}\vec{\bf{r}} \, .
\end{split}                              
\label{DENF-1}
\end{equation}
Here, $A$ is the nucleon number and $N$ is the power for the moment.

The reduced moments $\langle \rho^{N} \rangle/\langle \rho\rangle^{N}$ for our simulations are presented in Figs.~\ref{fig:fig7} and \ref{fig:fig8}.   Fig.~\ref{fig:fig7} shows the moments at the average density of $0.2 \, \rho_0$, as function of temperature for different $N$ order. The moments rise when temperature drops to $10 \, \text{MeV}$ or lower, with the rise in the moments becoming more pronounced with the rise in $N$, reflecting the clumping of the matter. Similar clumping signatures have been effectively employed in the simulations for the vicinity of the high-energy QCD phase transition~\cite{JS12,LF17}. Fig.~\ref{fig:fig8} displays  results for the moments both with and without MF at different average densities.  Without MF, the moments change little with temperature, no matter what average density, as no clumping occurs. With MF the enhancement in the moments occurs for the same temperatures and densities that the relative drop in shear viscosity occurs in Fig.~\ref{fig:fig5}, underscoring the connection of fragment formation to the change in viscosity.

\section{Conclusions}
\label{summary}

In summary, we examined the stress tensor for a system of nucleons governed by the ImQMD version of the nuclear molecular dynamics, either employing or not the mean field, within the thermodynamic region of the liquid-gas phase transition.  The nucleons were enclosed in a periodic box with boundary conditions representing constant shear throughout space.  Following the SLLOD algorithm, we extracted the shear viscosity coefficient for the system, after it reached a stationary equilibrium.  We found that the mean field has little impact on the coefficient outside of the phase transition.  However, for lower average densities and temperatures in the region of the transition, the shear viscosity coefficient drops significantly when the mean field is included.  The drop can be attributed to the formation of fragments, or long-range correlations, of which presence can be revealed, in the box configuration, with average values of the moments of density. Also it is found that Coulomb interaction as repulsive interaction among protons reduces shear viscosity in the low temperature region.

The shear viscosity can be tested in nuclear collisions through examinations of momentum transfer between different momentum regions \cite{barker19}.  The region of the liquid-gas phase transition \cite{Fu06} is crossed in nuclear collisions at tens of MeV/nucleon.\\

\begin{acknowledgments}
X.G.D. expresses his gratitude to the University of Chinese Academy of Sciences for the scholarship awarded in 2017 (Grant No. N201709) that made this work possible.  P.~Danielewicz acknowledges support from the U.S.~Department of Energy Office of Science under Grant No.\ DE-SC0019209.   This work further received partial support from the National Natural Science Foundation of China under Contract No.\ 11890714, No. 12147101, No. 11947217 and No. 11875323, the Key Research Program of the CAS under Grant No.\ XDB34030000, China Postdoctoral Science Foundation Grant No.\ 2019M661332 and  Postdoctoral Innovative Talent Program of China No.\ BX20200098.
\end{acknowledgments}

\end{CJK*}
\end{document}